\documentclass[twocolumn]{aastex63}

\usepackage{graphicx}	% Including figure files
\usepackage{amsmath}	% Advanced maths commands
\usepackage{amssymb}	% Extra maths symbols
\usepackage{float}
\usepackage{longtable}
\usepackage{wasysym}
\usepackage{hyperref}

\definecolor{darkviolet}{rgb}{0.58, 0.0, 0.83}

\newcommand{\Msun}{{\ M_{\odot}} }

\received{}
\revised{}
\accepted{}
\submitjournal{ApJL}

\shortauthors{Van \& Ivanova}
\graphicspath{{./}{Images/}}
\begin{document}

\title{Evolving LMXBs: CARB Magnetic Braking}

\correspondingauthor{Kenny X. Van}
\email{kvan@ualberta.ca}

\author[0000-0003-3862-5826]{Kenny X. Van}
\affiliation{Department of Physics, University of Alberta, Edmonton, AB, T6G 2E7, Canada}

\author[0000-0001-6251-5315]{Natalia Ivanova}
\affiliation{Department of Physics, University of Alberta, Edmonton, AB, T6G 2E7, Canada}



\begin{abstract}

The formation of low-mass X-ray binaries (LMXBs) is an ongoing challenge in stellar evolution. The important subset of LMXBs are the binary systems with a neutron star (NS) accretor. In NS LMXBs with non-degenerate donors, the mass transfer is mainly driven by magnetic braking.  The discrepancies between the observed mass transfer (MT) rates and the theoretical models were known for a while. Theory predictions of the MT rates are too weak and differ by an order of magnitude or more. Recently, we showed that with the standard magnetic braking, it is not possible to find progenitor binary systems such that they could reproduce -- at any time of their evolution -- most of the observed persistent NS LMXBs. In this {\it Letter} we present a modified magnetic braking prescription, CARB (Convection And Rotation Boosted). CARB magnetic braking combines two recent improvements in understanding stellar magnetic fields and magnetized winds --  the dependence of the magnetic field strength on the outer convective zone and the dependence of the Alfv\`en radius on the donor's rotation. Using this new magnetic braking prescription, we can reproduce the observed mass transfer rates at the detected mass ratio and orbital period for all well-observed to-the-date Galactic persistent NS LMXBs. For the systems where the effective temperature of the donor stars is known, theory agrees with observations as well.
\end{abstract}



\section{Introduction}
\label{Introduction}

Understanding the evolution of stars in binary systems relies heavily on the adopted laws of angular momentum loss which affect the change in orbital separation. One of the ways to lose angular momentum in a binary system is through magnetic braking (MB) \citep{Verbunt1981}. In this concept, the donor loses its angular momentum through a magnetized wind, and then, through tidal friction, replenishes the donor's angular momentum using the orbital angular momentum. MB is the dominant angular momentum loss mechanism in binaries wider than %about a day
a few hours
in orbital period, whereas gravitational radiation dominates in close binaries \citep{Rappaport1983}. More recently, circumbinary disks have been shown to effectively remove angular momentum and reproduce ultra compact binaries \citep{Ma2009b}. Unfortunately, circumbinary disks appear to be rare in LMXBs and there are significant uncertainties in the disk parameters \citep{Ma2009b}. Additionally, our work includes systems with wider periods that UCXB., As such, we will be focusing only on MB.

The choice of the adopted MB prescription has large overarching effects on the evolution of the binary -- stronger MB will shrink a binary faster, resulting in a higher mass transfer (MT) rate. The
most 
widely-used assumption in stellar simulations is %to employ 
the ``Skumanich'' MB \citep{Skumanich1972}; its application to binary systems is usually %used in 
the form provided in \cite{Rappaport1983}. The standard MB law, as well as some of its modifications, %has been shown to fail 
fails 
to reproduce the observed persistent NS LMXBs \citep{Van2019}. Examples of some modified MB schemes include those which focus on a subset of LMXBs such as Ap/Bp donors \citep{Justham2006}, or dampen the MB strength at high rotation rates \citep{Sills2000, Ivanova2003}.

Some advances in understanding the characteristics of the magnetized wind from a star were made recently. First, \cite{Reville2015} has included the effect of stellar rotation on the Alfv\`en radius. Secondly, %connecting the strength of the surface of the magnetic field with the convective turnover time 
the convective turnover time has been linked to the strength of the surface magnetic field 
\citep{Parker1971, Noyes1984, Ivanova2006}. 
In \S2, we derive the new CARB (Convection And Rotation Boosted) MB which takes into account both advances. In \S3, we use the new MB to evolve the grid of progenitor binaries, in a similar manner as  done in \cite{Van2019}. In \S4, we compare the results of the simulations with the observed persistent NS LMXBs. Finally, in \S5 we summarize our key results in this \textit{letter}.

\section{Magnetic Braking}
\label{sec:MB_derivation}

The loss of the angular momentum due to magnetic braking is derived following  steps similar to those outlined in \cite{Van2019}. 

First, we assume spherical symmetry, which results in the angular momentum lost being

\begin{equation}
    \Dot{J}_{\rm MB}=-\frac{2}{3}\Omega\dot{M}_{\rm W}R^2_{\rm A}.
\label{eq:mb_init}
\end{equation}

\noindent $\Dot{M}_{\rm W}$ denotes the wind mass loss rate, $\Omega$ is the rotation rate, and $R_A$ is the Alfv\`en radius. Assuming a radial magnetic field, 

\begin{equation}
    \left(\frac{R_A}{R}\right)^2=\frac{B_s^2 R^2}{4\pi R_A^2 \rho_A v_A^2}=\frac{B_s^2 R^2}{\Dot{M}_{\rm W} v_A}. \\
\label{eq:mb_reville_p1}
\end{equation}

\noindent Here $R$ is the radius of the star, $B_s$ is the surface magnetic field strength, $v_a$ is Alfv\`en velocity, and $\rho_a$ is the density of the wind at the Alfv\`en radius. Total mass loss with the wind is $\Dot{M}_{\rm W}= 4\pi R_A^2 \rho_A v_A$. The velocity of a normal stellar wind, when it reaches the Alfv\`en radius, can be found from energy conservation, and  expressed using the surface escape velocity $v_{\rm esc}$:

\begin{equation}
    \frac{v_A}{v_{\rm esc}}=\left(\frac{R}{R_A}\right)^{1/2}.
\label{eq:alfv_vel}
\end{equation}

\noindent In the case when the star and its attached magnetic field rotate, the regular stellar wind can also be additionally accelerated by the time it reaches the Alfv\`en radius. This acceleration was tested by \cite{Matt2012} and was shown to have a non-negligible effect. \cite{Reville2015} parametrized the additional acceleration by replacing the surface escape velocity with a modified velocity, which includes the effects of rotation. Using this variable instead in Equation \ref{eq:mb_reville_p1} gives us

\begin{equation}
    \left(\frac{R_A}{R}\right)^3=\frac{B_s^4 R^4}{\Dot{M}_{\rm W}^2}\times\frac{1}{v_{\rm esc}^2+2\Omega^2 R^2/K_2^2}, \\
\label{eq:alfv_r_reville2}
\end{equation}

\noindent where $K_2=0.07$ in this equation is a constant obtained from a grid of simulations by \cite{Reville2015}. $K_2$ sets the limit where the rotation rate begins to play a significant role. In this approach, the Alfv\`en radius shrinks as the rotation rate increases, weakening the angular momentum loss in fast rotating binaries. Plugging this form of the Alfv\`en radius into the angular momentum equation gives a new prescription for angular momentum loss,

\begin{equation}
    \Dot{J}_{\rm MB}=-\frac{2}{3}\Omega\dot{M}_{\rm W}^{-1/3} R^{14/3} B_s^{8/3}\left(v_{\rm esc}^2+2\Omega^2 R^2/K_2^2\right)^{-2/3}. \\
\end{equation}

\noindent Substituting %the magnetic field strength of the star with a scaling relation 
a convective turnover scaling relation for the magnetic field strength of the star 
\cite[see][for a discussion as to why this is justified]{Van2019}, we get the modified magnetic braking prescription used in our simulations,

\begin{equation}
\begin{split}
    \Dot{J}_{\rm MB}=&-\frac{2}{3}\dot{M}_{\rm W}^{-1/3} R^{14/3} \left( v_{\rm esc}^2 + 2 \Omega^2 R^2/K_2^2\right )^{-2/3}\\
    &\times\Omega_\odot\  B_{\odot}^{8/3}\ \left(\frac{\Omega }{\Omega_\odot}\right)^{11/3}\left(\frac{\tau_{\rm conv} }{\tau_{\odot, \rm conv}}\right)^{8/3}\ .
\end{split}
\end{equation}

\noindent The magnetic field strength on the surface of the Sun is on average $B_s=1 \rm\ G$ with a
rotation rate and 
convective turnover time of $\Omega_\odot \approx 3\times10^{-6} \ \rm s^{-1}$ and $\tau_{\odot, \rm conv}=2.8 \times 10^6 \ \rm s$, respectively. Both solar values used here were found using the same method from \cite{Van2019}. The value used for $\tau_{\odot, \rm conv}$ is similar to those found by \cite{Ma2009a} and \cite{Landin2010} of 28.4d and 38.2d respectively. While our value deviates slightly from those found in other works, what is important is that our calculations are self-similar between different stars: the method used to calculate the normalization factor and the turnover time of each of our simulated systems is the same.

\section{Evolution through the mass transfer}
\label{sec:results}

\begin{figure*}
    \centering
    \includegraphics[width=\columnwidth]{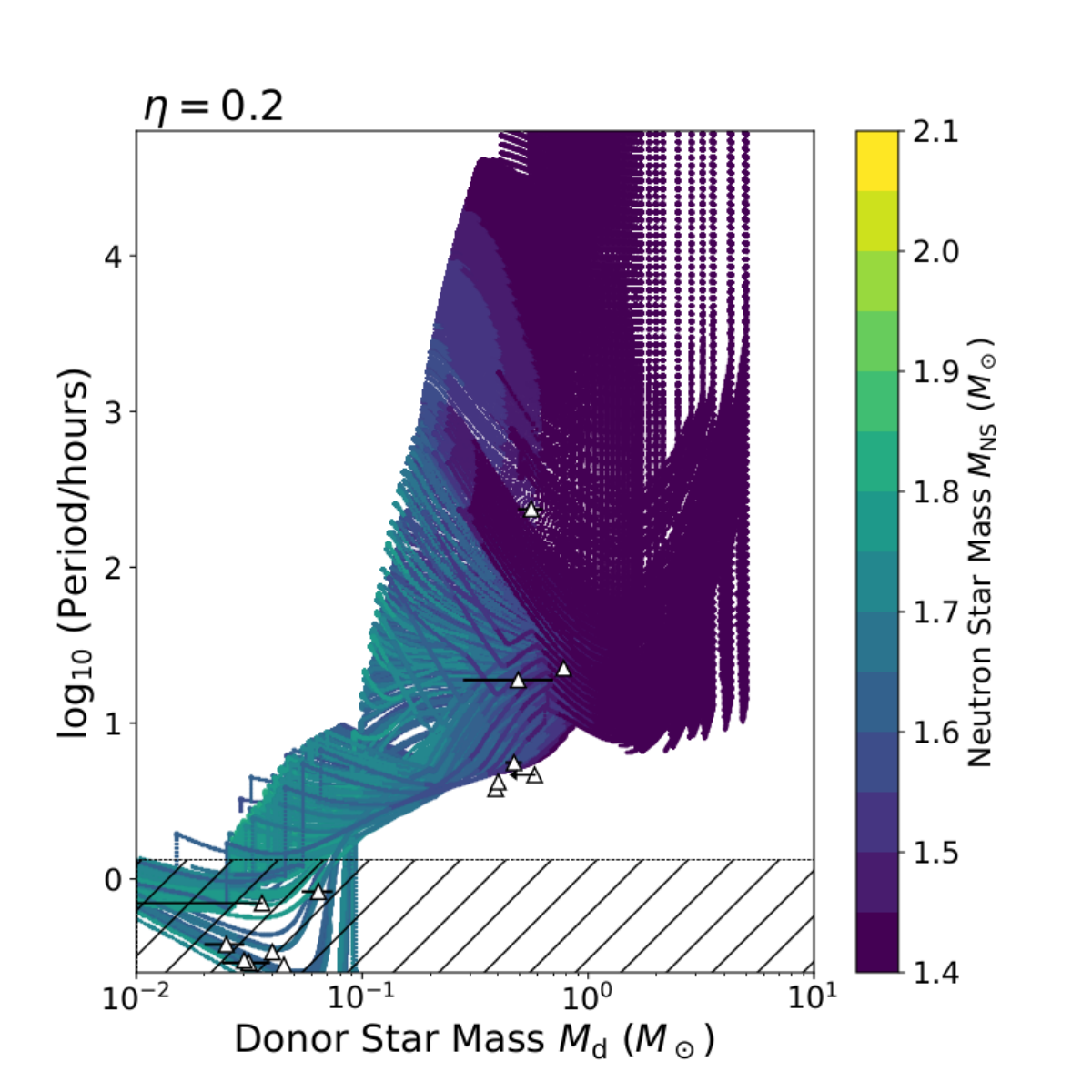}
    \includegraphics[width=\columnwidth]{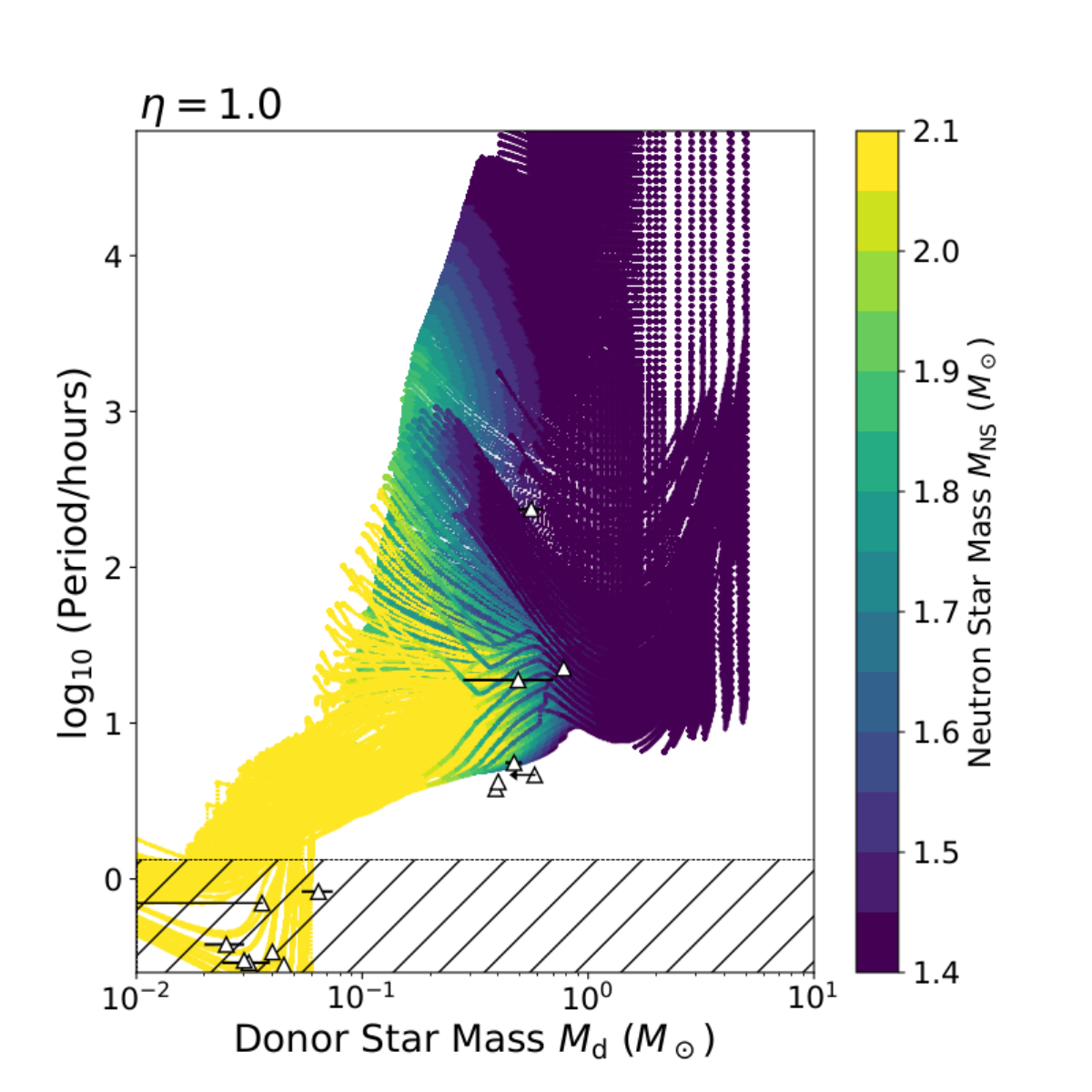}
    \caption{The evolution of $M_{\rm NS}$ during the mass transfer.  The $\eta$ value denotes the MT efficiency. The triangle symbols represent persistent LMXBs \citep[data from][]{Van2019}.}
    \label{fig:eta_comp}
\end{figure*}

We follow the method described in \cite{Van2019} and test the MB on  progenitor binaries seeded on a grid of periods and donor masses. The initial periods range from $-0.4 \leq \log_{10}(P/\rm day) \leq 4$ in steps of $\Delta \log_{10}(P)=0.05$.  The initial donor masses range from $1.0 \leq M_d/M_\odot \leq 7.0$ with a variable step size. The donor mass has steps of $\Delta M_d=0.1\Msun$ when $M_d \leq 2.4 M_\odot$, $\Delta M_d=0.2 M_\odot$ for $2.4 < M_d/M_\odot \leq 3$, $\Delta M_d=0.5 M_\odot$ when $3 < M_d/M_\odot \leq 5$ and $\Delta M_d=1.0 M_\odot$ for any initial donor mass exceeding $5 \Msun$. The stars have initial metallicity $Z=0.02$. All NSs start with a seed mass of $M_{\rm NS}=1.4M_\odot$. The chosen grid encompasses all binaries that could start the mass transfer at some point of their evolution.

To evolve the initial binaries, we use the stellar code \texttt{MESA}\footnote{\url{http://mesa.sourceforge.net}} (Modules for Experiments in Stellar Astrophysics) revision 11701 \citep{Paxton2011, Paxton2013, Paxton2015, Paxton2018, Paxton2019} and May 2019 release of  \texttt{MESASDK}\footnote{\url{http://www.astro.wisc.edu/~townsend/static.php?ref=mesasdk}}\footnote{The modifications to MESA to include modified MB will be available on the MESA marketplace.}
. 

Here, we refine the method described in \cite{Van2019} by taking into account the efficiency of the mass transfer. The rate of the mass gain of the NS $\dot{M}_{\rm NS}$ is proportional to the rate of the mass accretion $\dot{M}_{\rm acc}$, but is less than that due to conversion of some accreted mass into gravitational binding energy: 

\begin{equation}
\begin{split}
    \dot{M}_{\rm NS}=\dot{M}_{\rm acc}f_{\rm BE} \ .
\end{split}
\end{equation}

\noindent Here $f_{\rm BE}$ is the so-called binding energy factor. Depending on the equation of the state of the NS, $f_{\rm BE} \approx 0.85 - 0.90$ \citep{Lattimer2007}. Some fraction of the material accreted onto the NS will be converted to gravitational binding energy and is controlled by $f_{\rm BE}$.

In addition, not all mass transferred through $L_1$ has to be accreted by the NS --  it may be reduced by a number of effects, for example, the propeller effect is a mechanism where the magnetic field deflects away accreting material \citep{Romanova2018}. Indirect evidence for the accretion inefficiency comes from observations of millisecond pulsars. 
If the accretion rate was the same as the mass transfer rate $\dot{M}_{\rm tr}$, many of these binaries are expected to contain high mass neutron stars.  
However, the observations do not support this \citep{Antoniadis2012, Antoniadis2016}. 
An analytic description of the efficiency of mass transfer is not currently known. 
\cite{Antoniadis2012} %found through observations 
calculated that accretion onto 
 the pulsar PSR J1738+0333 had an efficiency $\epsilon \sim 0.1 - 0.3$, while a more recent statistical study looking at a number of pulsars estimated that their accretion efficiency was between $\epsilon \sim 0.05 - 0.2$ \citep{Antoniadis2016}. 
We will combine the efficiency and the binding energy factor into one value $\eta$. The material accreted by the NS is less than that transferred, 

\begin{equation}
\begin{split}
    \dot{M}_{\rm acc}=\eta\dot{M}_{\rm tr} \ .  \\
    % \dot{M}_{\rm NS} &= \textrm{min}\left(\eta \dot{M}_{\rm tr}, \dot{M}_{\rm edd} \right)\ .
\label{eq:mt_eff}
\end{split}
\end{equation}

In Figure \ref{fig:eta_comp} we demonstrate how the choice of $\eta$ affects the mass of the final NSs.
With $\eta=1$, NSs in most systems become more massive than $2M_\odot$, once $M_{\rm d}<0.4M_\odot$. 
While $M_{\rm NS}$ is predicted to extend up to 
$\approx 2.1-2.2 M_\odot$
%$\approx 2.16 M_\odot$ \citep{Ozel2016}, 
%only two observed NSs have their inferred mass above $2.0 M_\odot$
no NSs have accurately and reliably measured masses exceeding $2.0 M_\odot$ 
\citep{Antoniadis2013, Thankful2019, Rezzolla2018}. 
The rarity of high mass NSs appears to contradict our results when assuming high efficiency. 
With $\eta=0.2$, the maximum mass of the NS is of order $M_{\rm NS} \sim 1.8M_\odot$. 
This value is within the 
%reasonable 
range of $1.1 \lesssim M_{\rm NS}/M_\odot \lesssim 2$ for observed NSs \citep{Ozel2016}.
For our study in this {\it Letter}, we therefore adopt $\eta=0.2$.

The efficiency factor will have a variety of effects on the binary system. The increased mass ejected from the system will increase the amount of angular momentum lost and limit how quickly $M_{\rm NS}$ grows. The efficiency controls how much material is accreted onto the compact object, which sets the luminosity of the system. The mass transfer efficiency is not constant throughout the entire evolution, and as a rough approximation we will estimate that the luminosity of our system can be approximated by $L=0.6 G\dot{M}_{\rm tr} M_{\rm NS}/R_{\rm NS}$. We increase the size of our MT bins used in the analysis to compensate for the uncertainty in MT efficiency. If the $\eta$ parameter used here were applied to the results from \cite{Van2019}, the overall ability of a system to reproduce an observed LMXB would remain unchanged or decrease as the MT rate required to explain the observed X-ray luminosity may be increased.

\section{Comparison with the observed population of LMXBs}
\label{sec:stat}

\begin{figure}
    \centering
    \includegraphics[width=\columnwidth]{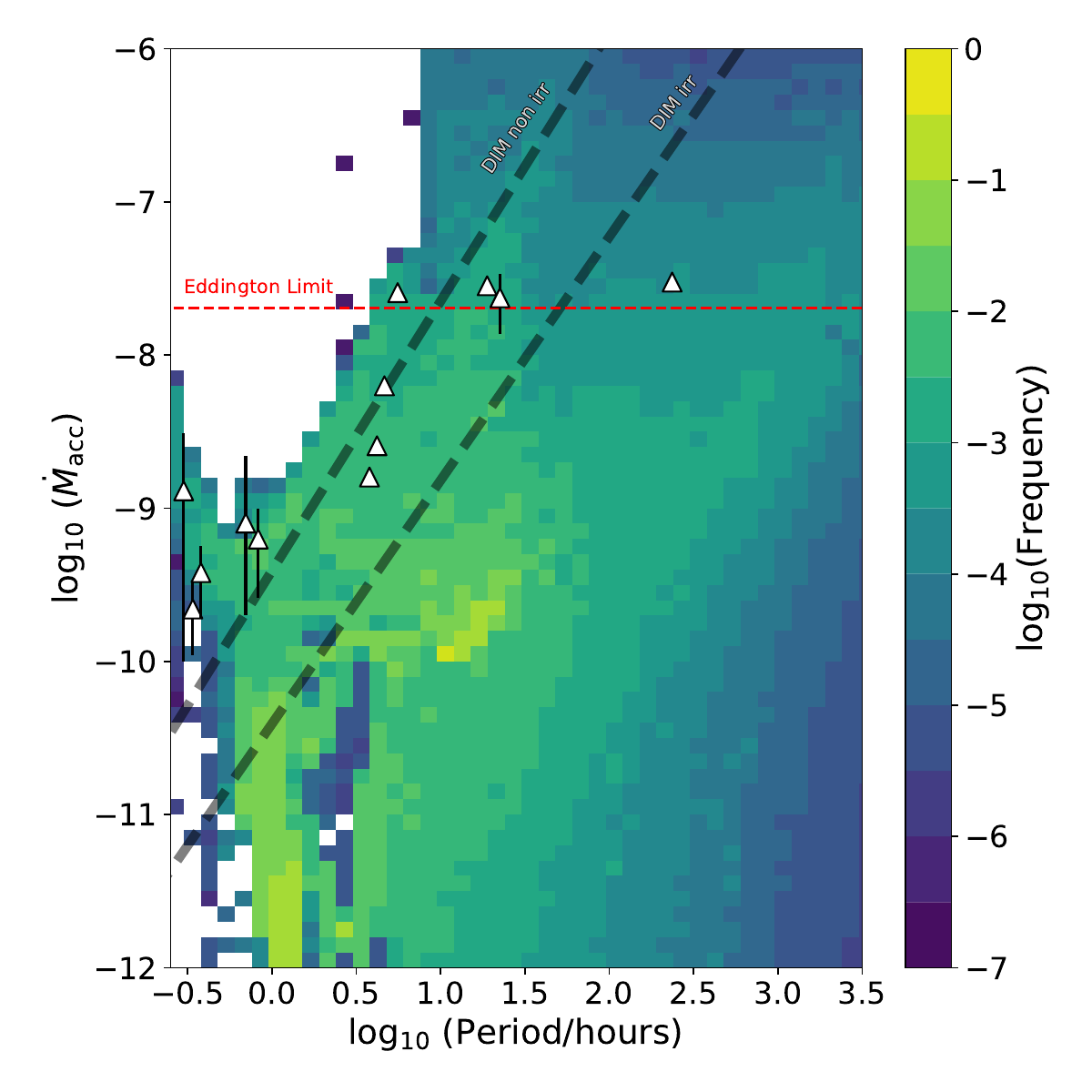}
    \caption{The relative probability of finding a system in a given bin in period-MT space. Each bin spans a width and height of 0.1 in $\log_{10}(P)$ and $\log_{10}(\dot{M}_{\rm  acc})$. The symbols used are the same as in figure \ref{fig:eta_comp}. The two grey dashed lines represent the critical MT separating persistent and transient systems for $M_{\rm NS}=1.4\rm M_\odot$ as described by the disk instability model (DIM) \citep{Coriat2012}. The upper line includes the effects of irradiation while the lower line does not.}
    \label{fig:density}
\end{figure}

It has been shown that the results of the MT simulations can be misleading in determining the legitimacy of adopted MB prescriptions if only two parameters are compared between simulated and observed systems \citep{Pavlovskii2016}. At least three parameters -- for example, the period, the MT rate and the mass ratio -- are necessary for determining if a given MB prescription is effective. The effective temperature of the donor could also play a significant role in discriminating the adopted MB laws \citep{Justham2006}.

It is hard to visualize the compatibility of three or more parameters in the same figure.
In Figure~\ref{fig:density}, we show the maximum relative probability for any of the simulated MT systems to have a specific MT rate and orbital period, as well as the MT rates and orbital periods of observed persistent NS LMXBs \citep[data is taken from][]{Van2019}.
This relative probability, or frequency, is calculated using the following steps:

\begin{enumerate}
\item $\tau^{mn}_{\rm tot}$ is the total evolutionary time of a binary system for an initial mass $m$ and initial period $n$.

\item $\tau_{ij}^{mn}$ is the amount of time the initial $m, n$ binary spends in an observed $i, j$ period and MT bin.

\item $f_{ij}^{mn}=\tau_{ij}^{mn}/\tau^{mn}_{\rm tot}$ is the frequency with which a given combination of mass and period appears in an observed bin of interest.

\item $f_{ij}=$ max $(f_{ij}^{mn})$ is the maximum frequency from all the simulated binaries, and is plotted in Figure \ref{fig:density}.
\end{enumerate}

Within this period-MT parameter space, all of the observed persistent NS LMXBs appear to be reproducible by the simulated MT systems. This apparent match does not guarantee that the simulated systems will reproduce the observed systems when additional parameters are included.

Let us briefly describe the methodology for the comparison in 3-parameter space \citep[for details, see][]{Van2019}.
Each observed system is assigned a 3-dimensional cuboid, where the cuboid is roughly centred in the observed properties. The size of of the cuboid in period is $\delta \log_{10}P=0.05$ and the size of the cuboid in mass ratio and MT rate depend on the uncertainty with which the observed value was determined, see Table \ref{table:combined_table}.

\begin{table*}
\footnotesize
\caption{\textbf{Binned Properties of LMXBs}}
\centering
\begin{tabular}{l | llllll}
% \hline

System Name          & $\log_{10}(P/\rm hr)$  & q            & $\log_{10}(\dot M_a)$ & $\tau_{\rm max}$ (years) & $A_{\rm sys}/A_{\rm tot}$ & $f_{\rm LMXB} $\\
\hline
4U 0513-40           & [-0.57, -0.52]  & [0.01, 0.06] & [-9.0, -8.4]          & $ 5.87\times10^{6}     $ & $ 1.72\times10^{-3} $     & $ 4.38\times10^{-2} $\\ 
2S 0918-549          & [-0.56, -0.51]  & [0.01, 0.06] & [-9.6, -8.4]          & $ 5.63\times10^{6}     $ & $ 1.72\times10^{-3} $     & $ 4.38\times10^{-2} $\\ 
4U 1543-624          & [-0.54, -0.49]  & [0.01, 0.06] & [-8.9, -8.4]          & $ 5.85\times10^{6}     $ & $ 1.54\times10^{-3} $     & $ 4.38\times10^{-2} $\\ 
4U 1850-087          & [-0.48, -0.43]  & [0.01, 0.06] & [-9.8, -8.2]          & $ 1.58\times10^{7}     $ & $ 2.92\times10^{-3} $     & $ 8.82\times10^{-2} $\\ 
M15 X-2              & [-0.44, -0.39]  & [0.01, 0.06] & [-9.5, -8.9]          & $ 2.43\times10^{7}     $ & $ 2.92\times10^{-3} $     & $ 5.37\times10^{-2} $\\ 
4U 1626-67           & [-0.17, -0.12]  & [0.01, 0.06] & [-9.5, -8.4]          & $ 7.39\times10^{7}     $ & $ 2.92\times10^{-3} $     & $ 1.05\times10^{-1} $\\ 
4U 1916-053          & [-0.10, -0.05]  & [0.03, 0.08] & [-9.4, -8.7]          & $ 6.14\times10^{7}     $ & $ 1.03\times10^{-3} $     & $ 8.86\times10^{-2} $\\ 
4U 1636-536          & [0.56, 0.61]    & [0.15, 0.40] & [-8.9, -8.4]          & $ 2.32\times10^{7}     $ & $ 5.49\times10^{-3} $     & $ 5.85\times10^{-2} $\\  
GX 9+9               & [0.60, 0.65]    & [0.20, 0.33] & [-8.5, -8.0]          & $ 1.39\times10^{7}     $ & $ 4.46\times10^{-3} $     & $ 9.11\times10^{-2} $\\  
4U 1735-444          & [0.65, 0.70]    & [0.29, 0.48] & [-8.2, -7.7]          & $ 1.11\times10^{7}     $ & $ 4.97\times10^{-3} $     & $ 1.44\times10^{-2} $\\  
2A 1822-371          & [0.73, 0.78]    & [0.26, 0.36] & [-7.6, -7.1]          & $ 5.95\times10^{6}     $ & $ 6.69\times10^{-3} $     & $ 7.06\times10^{-2} $\\  
Sco X-1              & [1.26, 1.31]    & [0.15, 0.58] & [-7.8, -7.1]          & $ 5.42\times10^{6}     $ & $ 1.20\times10^{-3} $     & $ 4.32\times10^{-3} $\\  
GX 349+2             & [1.33, 1.38]    & [0.39, 0.65] & [-7.8, -7.1]          & $ 1.21\times10^{7}     $ & $ 4.46\times10^{-3} $     & $ 4.25\times10^{-3} $\\  
Cyg X-2              & [2.35, 2.40]    & [0.25, 0.53] & [-7.5, -7.0]          & $ 7.99\times10^{4}     $ & $ 1.72\times10^{-3} $     & $ 6.65\times10^{-4} $\\  
\hline

\end{tabular}
\label{table:combined_table}
\begin{flushleft}
\textbf{Notes.} The binned properties of observed persistent NS LMXBs taken from \cite{Van2019}. 
This table is adapted from Table 4 from \cite{Van2019}. Again the periods are in hours and the mass accretion rate $\dot M_a$ is in $M_\odot\ yr^{-1}$. The bin ranges were chosen to span the errors in the given observed property with the bins centred on the observed values. $\tau_{\rm max}$ is the maximum amount of time a given simulated system spends in the observed bin of interest. $A_{\rm sys}/A_{\rm tot}$ is the fraction of our tested parameter space that can reproduce the system of interest. These two quantities give an indication to how long a simulation appears similar to an observed LMXB and how many systems could reproduce these properties.
\end{flushleft}
\end{table*}

We can find the maximum time that an individual simulation spends in a bin of interest, $\tau_{\rm max}$, and what fraction of their MT evolution they spend in the given bin, $f_{\rm LMXB}$. We also can find the fractional area of the initial parameter space that reproduces the binary $A_{\rm sys} / A_{\rm tot}$. These three numbers can indicate how plausible it is to produce the observed NS LMXBs. 
The value of 
$\tau_{\rm max}$  indicates how long a system can remain in this state,
and thus how likely it is to be detected. 
$A_{\rm sys} / A_{\rm tot}$ shows how stringent the initial parameter space is for reproducing a given LMXB. 
A larger $A_{\rm sys} / A_{\rm tot}$ implies that many systems can reproduce an observed system. 
$A_{\rm tot}$ spans our entire parameter space of seed masses and periods. In our case $A_{\rm tot}=29.1475$. For example, we find that Cyg X-2 only has 2 progenitor systems, these two progenitor systems span a total area of $A_{\rm sys}=0.05$  which results in $A_{\rm sys}/A_{\rm tot}=1.72\times10^{-3}$.

As has been shown by \cite{Van2019}, once the constraint on the mass ratio is added, none of the previously used MB prescriptions can produce all of the observed persistent NS LMXBs, despite considering all possible initial binaries. For the non-reproducible systems, $A_{\rm sys}/A_{\rm tot}=0$. In Table~\ref{table:combined_table} we present the results for the CARB MB prescription. It is fascinating that with the modified MB prescription, all persistent LMXBs can be reproduced. 

We can further constrain the progenitors by looking at the effective temperature of the donor star. Determining the temperature of the companion is difficult, and this value is not known for most observed LMXBs. The systems where the donor's spectral type have been measured tend to be the widest LMXBs; Sco X-1, GX 349+2 and Cyg X-2. This additional fourth observed parameter will provide additional constraints to the progenitor mass and period combinations that result in binaries that can match all observed properties.

Sco X-1 was found to have a donor star that was later than K4 \citep{Sanchez2015}. This gives an approximate upper limit to the donor temperature to be $\lesssim 4800$ K. By matching our three previous properties of interest -- period, mass ratio and MT -- while constraining the donor temperature, we can further limit systems that reproduce Sco X-1. An example progenitor of Sco X-1 has a $1.1\Msun$ donor with an initial period of 2.82 days. This system simultaneously matches the period, mass ratio, MT and effective temperature of Sco X-1. When this progenitor evolves to the observed mass ratio and period, the MT rate and effective temperature of the binary are $2.3 \times 10^{8} \ M_\odot \rm\ yr^{-1}$ and $4685 \rm\ K$ respectively.

Cyg X-2 was found by \cite{Cowley1979} to have an A5-F2 donor star. A5-F2 spectral type stars have an approximate temperature range of 7000 - 8500 K. When comparing this to our MT systems, we find that the only progenitors that reproduce Cyg X-2 are binaries with an initial period between $P \approx 2.24 - 2.51 \rm\ days$ and an initial donor mass of $M=3.5 \Msun$. The mass transfer rates and effective temperatures of the $2.24 \rm\ day$ progenitor are $2.9 \times 10^{-8} \ M_\odot \rm\ yr^{-1}$ and $7265 \rm\ K$.

GX 349+2 is a system where the spectral class of the donor is given, but the literature related to this property is not in agreement. \cite{Penninx1991} found the donor of GX 349+2 to be a G5-M2 giant whereas \cite{Wachter1996} finds the donor could be a B2 main sequence donor. Our simulated results have a temperature ranging from $\approx 4800 - 5500$ K which correspond to a  K3-G5 donor star. An example progenitor of GX 349+2 is a binary with an initial donor mass of $M=1.1 \Msun$ and a seed period of $3.98 \rm\ days$. This progenitor has a MT rate of $8.2\times10^{-8} \ M_\odot \rm\ yr^{-1}$ and an effective temperature of $4845 \rm\ K$.

\section{Conclusions}
\label{sec:Conclusion}

We revised the MB prescription to include the effect of the donor's rotation on the wind's velocity, following \cite{Matt2012} and \cite{Reville2015}, as well as the effects of the donor's convective eddy turnover timescale and the donor's rotation on the generation of the surface magnetic field, following \cite{Parker1971, Noyes1984, Ivanova2006, Van2019}.

The new CARB MB prescription was applied to test the evolution of all binaries with a NS and non-degenerate donors that could experience the mass transfer at some point in their evolution.
The modelled MT systems were compared to the observed persistent NS LMXBs. Our simulations were required to match with %most 
observations in three parameters -- the MT rate, the orbital period and the mass ratio, with the effective temperature being used as a fourth parameter in select binaries.
Previously, it has been shown that the most commonly used MB prescription, also known as Skumanich MB \citep{Rappaport1983}, 
is not capable of reproducing most of the persistent NS LMXBs with orbital periods larger than about an hour.
With our modified MB, we can reproduce all observed persistent NS LMXBs. 

We note that the ``Intermediate'' prescription considered in \cite{Van2019} reproduced all of the LMXBs of interest as well, 
although that
description was not explicitly derived -- it was created by adding  ad-hoc wind boosting and ad-hoc convection boosting. 
Both of these factors are taken into account in a more physical way in the  modified MB prescription presented here. Additionally, once the effective temperature is accounted for with the ``intermediate'' prescription, Sco X-1 could no longer be reproduced. The number of possible  progenitors of Cyg X-2 also significantly drops, to only one system. 

Our simulations do not include additional effects such as irradiation, or atypically strong magnetic fields similar to those found in Ap stars. While these effects might be invoked to explain a specific individual system, they could not be used to explain the evolution of the entire population of MT binaries. The inclusion of rotational effects on the Alfv\`en radius, and magnetic field dependence on convective turnover time, resulted in CARB MB being able to reproduce all of the observed persistent NS LMXBs. 
We unequivocally 
recommend the use of the CARB MB prescription instead of the Skumanich MB, to model both Galactic and  extragalactic NS LMXBs.

Once the governing angular momentum loss law is constrained, our next step will be to recover and constrain the properties of the plausible progenitor systems, and the required formation rates of these progenitors to produce the observed numbers of LMXBs. We also intend to expand our sample size to include BHs and any additional well constrained NSs available. This will be a topic of our future research.

\acknowledgments

We would like the thank the referee for helpful comments.
N.I. acknowledges that a part of this work was
performed at the KITP, which is supported in part by the
National Science Foundation under Grant No. NSF PHY-1748958. 
N.I. acknowledges support from CRC program and
funding from NSERC Discovery under Grant No. NSERC RGPIN-2019-04277.
K.V. acknowledges the helpful comments from Craig Heinke.

\facility{ComputeCanada.}


\vspace{5mm}
% \facilities{facilities}


% \software{Software}



% \appendix




\end{document}